\documentclass[a4paper,12pt]{amsart}

\usepackage{graphicx, amsmath,setspace}

\begin{document}
\title{A note on the CAPM with endogenously consistent market returns}
\author{Andreas Krause}
\address{University of Bath, Department of Economics, Claverton Down, Bath BA2 7AY, Great Britain}
\email{mnsak@bath.ac.uk}
\begin{abstract}
 \noindent I demonstrate that with the market return determined by the equilibrium returns of the CAPM, expected returns of an asset are affected by the risks of all assets jointly. Another implication is that the range of feasible market returns will be limited and dependent on the distribution of weights in the market portfolio. A large and well diversified market with no dominating asset will only return zero while a market dominated by a small number of assets will only return the risk-free rate. In the limiting case of atomistic assets, we recover the properties of the standard CAPM.
\end{abstract}

\begingroup
\maketitle
\endgroup
\onehalfspacing
\section{Introduction}
The Capital Asset Pricing Model (CAPM) was developed by \cite{Sharpe64, Lintner65, Mossin66} from portfolio theory as introduced by \cite{Markowitz52}. It uses the idea that as the so-called optimal risky portfolio is independent of preferences, it will be held by all investors. Thus if this portfolio is held by all investors, it must be equivalent to the market portfolio. The resulting equilibrium results then in the relationship between expected asset returns and their risks, measured by the covariance of the asset with the market, commonly referred to as 'beta'. This relationship is one of the most widely used concepts in finance with applications ranging from evaluating investment opportunities and valuing assets, to explaining the cross section of asset returns.

In using the CAPM, it is commonly assumed, mostly implicitly, that the return on the market portfolio is exogenously given. However, as I argue in this note, this is generally not justified in an equilibrium model. Even though preferences would determine the expected market return, this market return needs to be consistent with the returns of the individual assets. While it might be appropriate to neglect this apect if we only consider a small number of assets that make up a negligible part of the market, the situation changes of we consider the market in its entirety to obtain equilibrium returns.

\section{Main result}
We have the standard CAPM as
\begin{equation}\label{eqn1}
  \boldsymbol\mu -r\boldsymbol\iota=\boldsymbol\beta\left(\mu_M-r\right),
\end{equation}
where $\boldsymbol\mu=\left[\mu_1, \mu_2, \ldots, \mu_N\right]^T$ denotes the vector of expected returns of the $N$ assets, $r$ the risk-free rate, $\boldsymbol\iota=\left[1, 1, \ldots, 1\right]^T$ a vector of 1s, $\boldsymbol\beta=\left[\beta_1, \beta_2,\ldots, \beta_N\right]^T$ the vector of betas for each asset, and $\mu_M$ the expected market return. With $\boldsymbol\omega=\left[\omega_1, \omega_2,\ldots, \omega_N\right]^T$ denoting the vector of weights for each asset in the market portfolio, we obviously have $\mu_M=\boldsymbol\omega^T\boldsymbol\mu$. Inserting this relationship into the CAPM of equation (\ref{eqn1}), we can transform the CAPM into $\left(\boldsymbol I-\boldsymbol\beta\boldsymbol\omega^T\right)\boldsymbol\mu=\left(\boldsymbol\iota-\boldsymbol\beta\right)r$, with $\boldsymbol I$ denoting the identity matrix. As $\text{det}\left(\boldsymbol I-\boldsymbol\beta\boldsymbol\omega^T\right)=1-\boldsymbol\beta^T\boldsymbol\omega=0$ due to the fact that $\boldsymbol\beta^T\boldsymbol\omega=1$, the matrix on the left hand side is not invertible. To solve for $\boldsymbol\mu$ we employ the Moore-Penrose Inverse, which for a matrix $\boldsymbol D$ is defined as $\boldsymbol D^+=\boldsymbol D^T\left(\boldsymbol D\boldsymbol D^T\right)^{-1}$. We eliminate rows of the matrix $\boldsymbol I-\boldsymbol\beta\boldsymbol\omega^T$ as needed to ensure the existence of the Moor-Penrose Inverse, in general only one row needs to be removed, assuming there are no identical assets in the market portfolio. Thus solving for the expected returns of the individual assets, we obtain
\begin{equation}\label{eqn5}
  \boldsymbol\mu=\left(\boldsymbol I-\boldsymbol\beta\boldsymbol\omega^T\right)^+\left(\boldsymbol\iota-\boldsymbol\beta\right)r.
\end{equation}

Unlike in the case of the traditional CAPM in equation (\ref{eqn1}), it is apparent equation (\ref{eqn5}) that the expected return of an asset does not only depend on its own risk, $\beta_i$, but the risks of all assets jointly.

Defining $\boldsymbol D=\boldsymbol I-\boldsymbol\beta\boldsymbol\omega^T$ for notational simplicity, we can derive the marginal effect of changing the risk $\boldsymbol\beta$ on the expected return $\boldsymbol\mu$ to be given as
\begin{eqnarray}\label{eqn6}
  \frac{\partial\boldsymbol\mu}{\partial\boldsymbol\beta}&=&\left(\left(\boldsymbol\omega^T\boldsymbol D^+\left(\boldsymbol\iota-\boldsymbol\beta\right)-1\right)\boldsymbol D^T+\left(\boldsymbol D^+\boldsymbol D-\boldsymbol I\right)\boldsymbol\omega\left(\boldsymbol\iota-\boldsymbol\beta\right)^T\right)\left(\boldsymbol D\boldsymbol D^T\right)^{-1}r\\\nonumber
  &=&\left(\mu_M-r\right)\boldsymbol D^++\left(\boldsymbol D^+\boldsymbol D-\boldsymbol I\right)\boldsymbol\omega\boldsymbol\mu^T\boldsymbol D^+.
\end{eqnarray}
We see that a change in the risk of one asset, $\beta_i$, does not only affect the expected return of that asset, $\mu_i$, but also the expected returns of other assets due to this matrix not being diagonal. This is in contrast to the standard CAPM, from which we obtain by differentiating equation (\ref{eqn1}) that
\begin{equation}\label{eqn7}
  \frac{\partial\boldsymbol\mu}{\partial\boldsymbol\beta}=\left(\mu_M-r\right)\boldsymbol I
\end{equation}
and thus only the expected return of the asset whose risk is changing, is affected.

\section{Limiting cases}
In our case with the market return being determined endogenously through the expected returns of its constituent assets, we can recover such a relationship in the limiting case where the influence of each asset on the market is negligible. To see this assume that $\forall i\in\left\{1,\ldots,N\right\}: \lim_{N\rightarrow\infty}\omega_i=0$, while maintaining $\boldsymbol\iota^T\boldsymbol\omega=1$, i.~e. the weights of all assets converge to zero as we increase the number of assets under consideration. In this case it is obvious that $\lim_{N\rightarrow\infty}\boldsymbol D=\boldsymbol I$, such that we recover equation (\ref{eqn7}) in the limit as $N\rightarrow\infty$. We thus see that if assets are atomistic, a change in the risk of an asset does only affect this asset itself. Hence, in this limiting case, our two results are consistent. Furthermore, as $\lim_{N\rightarrow\infty}\boldsymbol\mu=\left(\boldsymbol\iota-\boldsymbol\beta\right)r$, we find $\lim_{N\rightarrow\infty}\mu_M=\lim_{N\rightarrow\infty}\boldsymbol\omega^T\boldsymbol\mu=0$. In such a situation, the only possible expected market return is zero and, interestingly, stocks with higher $\beta_i$ would have lower expected returns.

In the more realistic case where at least some assets retain a substantial weight in a large and otherwise well diversified portfolio, any change of the risk of an asset would affect all assets. Another implication from equation (\ref{eqn5}) is that in the case of perfectly homogenous assets, such that $\boldsymbol\beta=\boldsymbol\iota$, we have $\boldsymbol\mu=\boldsymbol 0$ and all assets have expected returns of zero. Hence, if there is no heterogeneity in the risks of individual assets, the expected returns of all assets must equal zero, and subsequently also that of the market return. This is similar to the aforementioned cases of atomistic assets.

\begin{figure}[!t]
  \centering
  \includegraphics[width=\textwidth]{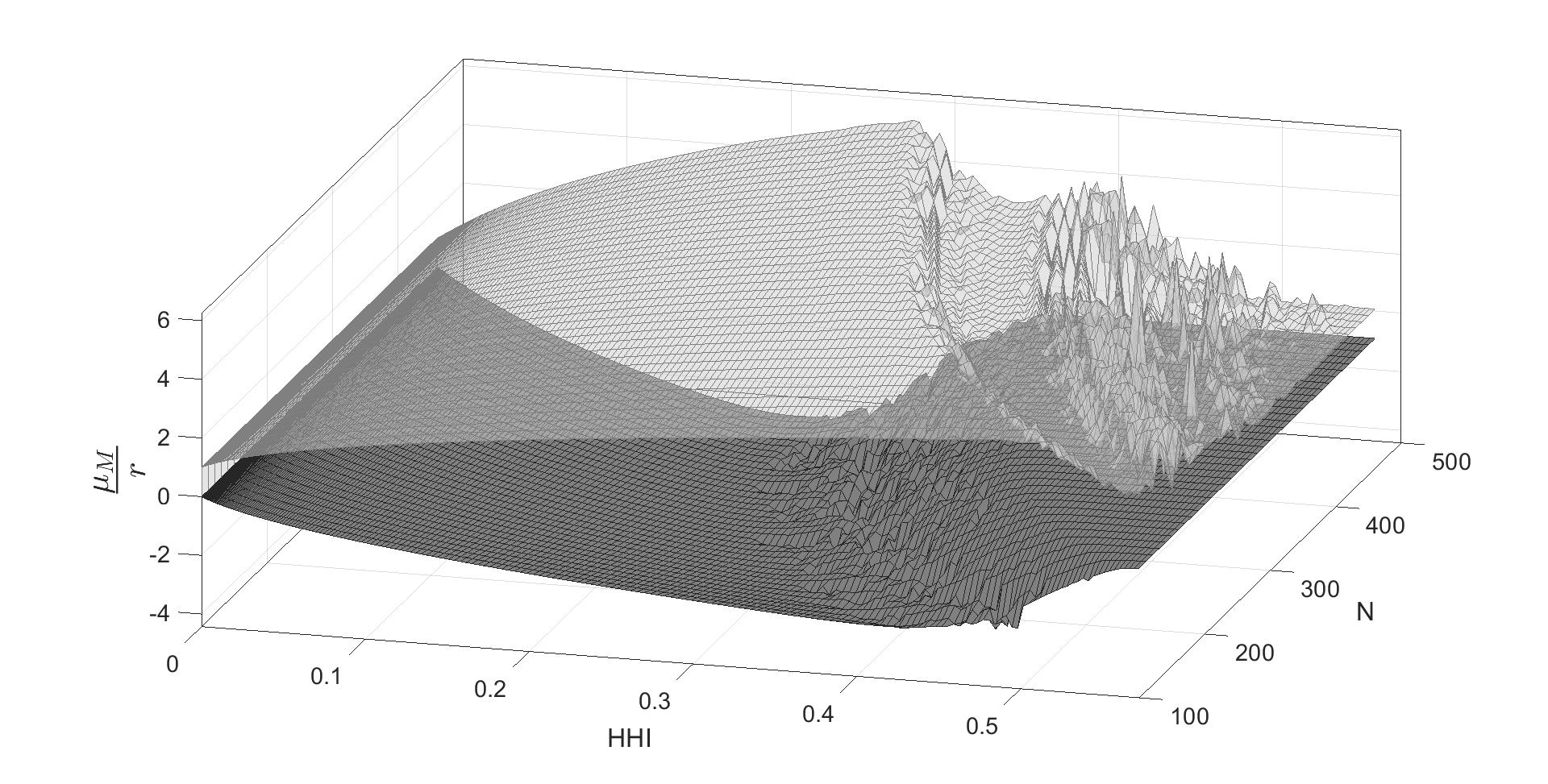}
  \caption{Supported range of expected market returns\newline \footnotesize{The graph shows the upper and lower limits of the supported expected market returns.}}\label{fig1}
\end{figure}

Furthermore, we can evaluate the market returns $\mu_M=\boldsymbol\omega^T\boldsymbol\mu$ for any given weights $\boldsymbol\omega$ or betas $\boldsymbol\beta$ that in equilibrium are feasible with the constraints that all weights must be non-negative and $\boldsymbol\iota^T\boldsymbol\omega=\boldsymbol\omega^T\boldsymbol\beta=1$. Using numerical methods, we find that only market returns within a certain range are feasible. This range will depend on the distribution of portfolio weights and the number of assets. Analysing this by assuming that $\omega_i\propto i^{-\gamma}$, where $\gamma\geq 0$ and $i\in\left\{1, \ldots, N\right\}$, we obtain how the maximal and minimal market returns possible will be affected. Using the normalized Herfindahl-Hirshman Index $HHI=\frac{\boldsymbol\omega^T\boldsymbol\omega-\frac{1}{N}}{1-\frac{1}{N}}$ to measure the concentration of weights in the market portfolio, we see from figure \ref{fig1} that in large markets for values of approximately 0.2-0.3, the supported market returns are covering the widest range from around $-4r$ to $6r$; for smaller markets higher asset concentrations are required to yield similar results. For markets that have a large number of approximately equally weighted assets (low $HHI$) or that are dominated by a small number of assets with high weights (high $HHI$), the maximal expected market return stays close to the risk-free rate.

In the extreme case that the $HHI$ is very high, e.g. only asset $i$ has a weight of 1 and all other assets a weight of 0 ($HHI=1$), we easily see that all but 1 row of the matrix $\boldsymbol D$ can be eliminated as those assets with a weight of 0 are irrelevant in the portfolio. Hence $\boldsymbol D=1-\beta_i$, giving rise to $\boldsymbol \mu=\left(\boldsymbol\iota-\boldsymbol\beta\right)\frac{r}{1-\beta_i}$, from which we get that $\mu_i=r$ and the return of the other assets being irrelevant as their weight is zero, which gives us $\mu_M=r$. If, on the other hand, all assets are equally weighted at $\omega_i=\frac{1}{N}$ and thus $HHI=0$, we have for any asset $i$ that $\boldsymbol D=\beta_i$ and therefore $\boldsymbol\mu=\left(\boldsymbol\iota-\boldsymbol\beta\right)\frac{r}{\beta_i}$, from which we easily get that $\mu_M=\boldsymbol\omega^T\boldsymbol\mu=0$, even though individual assets will have non-zero expected returns. It is thus only for sufficiently but not too heterogenous markets that wider ranges of market returns can be supported, especially returns exceeding the risk-free rate.

\section{Conclusions}
If the CAPM holds, in equilibrium the expected market return must be determined endogenously from the expected returns on individual assets. I have established that in this case the expected returns of asset $i$ is not only affected by their own risk through $\beta_i$, but also the risks of all other assets and a change in any risk affects all assets. Only in the limiting case where assets are atomistic in that there are an infinite number of assets and the weight of each assets becomes negligible, can we recover that the expected return of an asset is only affected by its own risk. In this case, however, the expected market return must be zero. In the general case where some assets retain a significant weight in the market portfolio a wider range of market returns can be supported, but the expected returns are determined jointly from the risk of all assets.

In reality most stock indices have a very low concentration of constituents, with the leading indices having values of the HHI in the region of 0.01. Such low values imply that the maximum expected market return will be limited to approximately 1.3 times the risk-free rate as can be inferred from figure \ref{fig1}. With historic returns on leading Western stock market indices in the region of 10\% p.~a. and risk-free rates of 6\% p.~a. it seems that the return of stocks market indices are too high to be consistent with this interpretation of the CAPM. It is, however, beyond the scope of this note to provide a detailed empirical analysis of the implication of these results and this this is left for suture research.

\bibliographystyle{unsrt}
\bibliography{CAPM}

\end{document}